\documentclass{achemso}
\usepackage{algorithm}
\usepackage{algpseudocode}
\usepackage{amssymb}
\usepackage{amsthm}
\usepackage{amsmath}
\usepackage{braket}
\usepackage{graphicx}
\usepackage{hyperref}
\usepackage{multirow}
\usepackage{pifont}
\usepackage{xcolor}

\setkeys{acs}{maxauthors=0}

\SectionNumbersOn

\usepackage[version=4]{mhchem}
\usepackage{tablefootnote}
\usepackage{float}
\usepackage{gensymb}
\usepackage{amsmath}

\usepackage{adjustbox} 

\author{Matthew Bousquet}
\email{bousquet@uchicago.edu}
\affiliation{Department of Chemistry, The University of Chicago, Chicago, Illinois 60637, USA}

\author{Jiawei Zhan}
\affiliation{Pritzker School of Molecular Engineering, The University of Chicago, Chicago, Illinois 60637, USA}

\author{Chunxin Luo}
\affiliation{Materials Science Division, Argonne National Laboratory, Lemont, Illinois 60439, USA}

\author{Alex B. Martinson}
\affiliation{Materials Science Division, Argonne National Laboratory, Lemont, Illinois 60439, USA}

\author{Francois Gygi}
\email{fgygi@ucdavis.edu}
\affiliation{Department of Computer Science, University of California Davis}

\author{Giulia Galli}
\email{gagalli@uchicago.edu}
\affiliation{Pritzker School of Molecular Engineering, The University of Chicago, Chicago, Illinois 60637, USA}
\alsoaffiliation{Department of Chemistry, The University of Chicago, Chicago, Illinois 60637, USA}
\alsoaffiliation{Materials Science Division, Argonne National Laboratory, Lemont, Illinois 60439, USA}

\title{Computational study of indium oxide photoelectrodes}

\keywords{XXX}

\begin{document}
\begin{abstract}
Using a combination of first principles molecular dynamics simulations (FPMD) and electronic structure calculations, we characterize the atomistic structure and vibrational properties of a photocatalytic surface of In$_2$O$_3$, a promising photoelectrode for the production of hydrogen peroxide.  We then investigate the surface in contact with water and show that the electronic states of In$_2$O$_3$ are appropriately positioned in energy to facilitate the two-electron water oxidation reaction (WOR) over the competing four-electron oxygen evolution reaction.  We further propose that the use of strained thin films interfaced with water is beneficial in decreasing the optical gap of In$_2$O$_3$ and thus utilizing a wider portion of the solar spectrum for the WOR.
\end{abstract}

\section{Introduction}

 Hydrogen peroxide is a highly effective oxidizing agent that sees wide use in industry \cite{Mavrikis2020}. For example, it is used in water treatment processes to remove organic impurities\cite{Farinelli2024}, through the formation of highly reactive hydroxyl radicals (•OH) which are used to degrade organic compounds, including aromatic and halogenated species \cite{Ma2021}.  Industrially, H$_2$O$_2$ is produced through the energy intensive anthraquinone process\cite{CamposMartin2006}, which involves catalytic hydrogenation, oxidation, distillation and purification. It is thus of interest to identify an alternative to produce H$_2$O$_2$, that avoids petrochemical solvents.

Hydrogen peroxide may be produced directly by splitting water, through the two electron water oxidation reaction (WOR): 
$2H_2O \rightarrow H_2O_2 +2H^+ + 2e^-$, that requires a thermodynamic potential of 1.77 V relative to the Standard Hydrogen Electrode (SHE).
Such reaction competes with the more favorable four electron oxygen evolution reaction (OER)\cite{Cook2010, Jones2024}: $2H_2O \rightarrow 2O_2 +4H^+ + 4e^-$, whose thermodynamic potential relative to SHE is 1.23 V. 


 A major challenge in the electrochemical production of hydrogen peroxide through the WOR is precisely to avoid the competing OER and also electrochemical degradation of peroxide, which requires only an overpotential of 0.67 V. Photoelectrochemical methods can, in principle, sidestep these issues\cite{Walter2010, Liu2019}, if photoanodes can be designed whose band edges are in a favorable position to trigger the WOR, instead of the OER.
In addition to band edges that straddle the WOR redox potential, an efficient photoanode \cite{Lewis2001} should be a water-stable n-type semiconductor so that the interfacial electric field generated by band bending drives the photoinduced holes toward the surface. Additionally, it should operate in a wide range of applied potentials and pH's, and possess high e$^-$/h$^+$ mobility and low resistivity\cite{Tan1994}, to limit the recombination of photoinduced e$^-$/h$^+$. 
 
The PEC WOR has only been realized on a handful of metal oxides, including TiO$_2$\cite{FUJISHIMA1972, Maurino2005}, SnO$_{2-x}$ \cite{Fan2013}, WO$_3$ \cite{Hill2012}, and BiVO$_4$ \cite{Nakabayashi2017}. 
Key figures of merit for several PEC WOR photoanodes are summarized in the SI (see Table S1).
So far,  BiVO$_4$ photoanodes have emerged as good candidates for PEC WOR, with the highest Faradaic Efficiency (FE, up to 90\% \cite{Baek2019}) as well as the highest reported photocurrents, around 2 mA cm$^{-2}$ at 1.23 V$_{RHE}$ \cite{Baek2019} for undoped BiVO$_4$.  However, such photocurrents are still small compared to those measured in several PEC OER systems \cite{Kay2006, Kim2013} and in general  bismuth vanadate has so far shown 
limited operational lifetimes. It is thus interesting to explore  alternative materials for use as photoanodes for WOR \cite{Mavrikis2021}.

 Here we focus on In$_2$O$_3$, which presents several attractive properties for PEC WOR, in spite of a large band gap. It 
 is water stable, has excellent free (Hall) carrier mobility, with values ranging from 10$^3$ cm$^2$/Vs to 10$^2$ cm$^2$/Vs at room temperature for single crystal films\cite{Bierwagen2010},  low resistivity (10$^{-2}$-10$^{-5},  \Omega$ cm \cite{Taggart2021} \cite{Ohta2000}) 
 and it is stable across a wide range of pH's and applied field strengths \cite{Pourbaix1974-ul}. While In$_2$O$_3$ has been used in photocatalytic applications, such as the degradation of organic pollutants \cite{Tang2023} and CO$_2$ reduction \cite{Wang2020, Deng2024}, it has not yet been actively explored for WOR.  

Early papers\cite{Miyauchi2002} reported moderate photooxidation properties of In$_2$O$_3$, possibly due to the exposure to water of (111) surfaces, later reported to be photo-inactive facets for water oxidation. 
In particular, Chu et al. \cite{Sun2013} showed that (111) and (001) surfaces show low and  high photocatalytic activity, respectively, and that nanocubes with only  (001) facets exposed to water had the highest PEC performance\cite{Meng2014}. 


Indium oxide nanocubes were initially grown with costly chemical vapor deposition methods, requiring temperatures of around 900 \degree C. However, recently, a bench-top synthesis was achieved\cite{Cho2019, Kim2020}, spurring new interest in In$_2$O$_3$ as a WOR platform. Experimentally, the stabilization of (001) facets depends on the growth conditions. 
 Notably, single (001) crystals grown by Hagleitner et al.\ using the flux method\cite{Hagleitner2012} exhibit wide terraces with stepped edges spaced 5 \r{A} apart, suggesting that there is one specific In layer preferably exposed at the surface, although the type of layer has not yet been determined in experiments. Several imaging (AFM, STM) and spectroscopic results (XPS) of samples grown in UHV  indicate a variety of surface reconstructions and terminations, including surface hydroxyls, peroxo species, and In-ad atoms \cite{Bourlange2009, Morales2009, KOROTCENKOV2005, Bierwagen2009}. (See Table S2 in the SI for a summary of  the surface sensitive characterization techniques  used in ultra-high-vacuum (UHV) for indium oxide).

Experiments on indium oxide/water interfaces have been conducted either on the (111) surface\cite{Chen2022, Wagner2017}, which is inactive in PEC \cite{Sun2013}, or on some commercial samples containing a mix of facets\cite{Donley2001, Brumbach2007}. From the experiments of Diebold et al \cite{Chen2022, Wagner2017}, it is known that the (111) surface is partially hydroxylated upon exposure to water.
Computational studies of In$_2$O$_3$/water interfaces have so far been limited. Zhou et al. \cite{Zhou2008} examined the binding of single water molecules to the polar (001) surface using Density Functional Theory (DFT) at the PBE level and found that they bind dissociatively on the surface. Agoston and Albe \cite{Agoston2011} investigated the thermodynamic stability of different low index In$_2$O$_3$ surfaces using again DFT at the PBE level, and found that the hydrogen terminated (hydroxylated) (001) surface has the lowest surface tension (i.e. it is the most stable) as a function of the water chemical potential.  However, no study to date has examined  the solvation of the surface, which naturally occurs during photoelectrochemical water oxidation processes. 


We note that while the exact mechanism of the PEC WOR on oxides is unknown\cite{Liu2019}, most proposed mechanisms\cite{Zhu2018, Viswanathan2015} include surface hydroxyls groups as key intermediates. Hydroxyl groups have been found to play an important role in the formation of H$_{2}$O$_2$ on several substrates, notably on electrified silica surfaces in contact with water microdroplets\cite{Chen2022-2}. Hence, understanding not only the solvation of In$_2$O$_3$ surfaces, but also the structure of hydroxyl groups on the surface and how water interacts with them, appears to be important to understanding the  photoelectrochemical activity of these surfaces. 

In this paper, we  report a first principles study of the dry hydroxylated (001) In$_2$O$_3$/surface and of the same surface in contact with water, to assess  its potential efficacy as a PEC WOR platform. 
 We focus on the (001) facet which, as discussed above, is photocatalytically active, and 
we investigate its electronic, structural, and vibrational properties. We address key questions about the interface with water, including which indium layer is most likely exposed at the surface, how the surface morphology is impacted by the presence of water, and how interactions with water affect the band alignment relative to the H$_2$O/H$_2$O$_2$ oxidation potential. 

The rest of the paper is organized as follows: the next section contains a description of the methods used in our study, followed by a discussion of our results and then our conclusions.

\section{Methods}
\subsection {Computational Methods}
\begin{figure}
  \includegraphics[width=\textwidth]{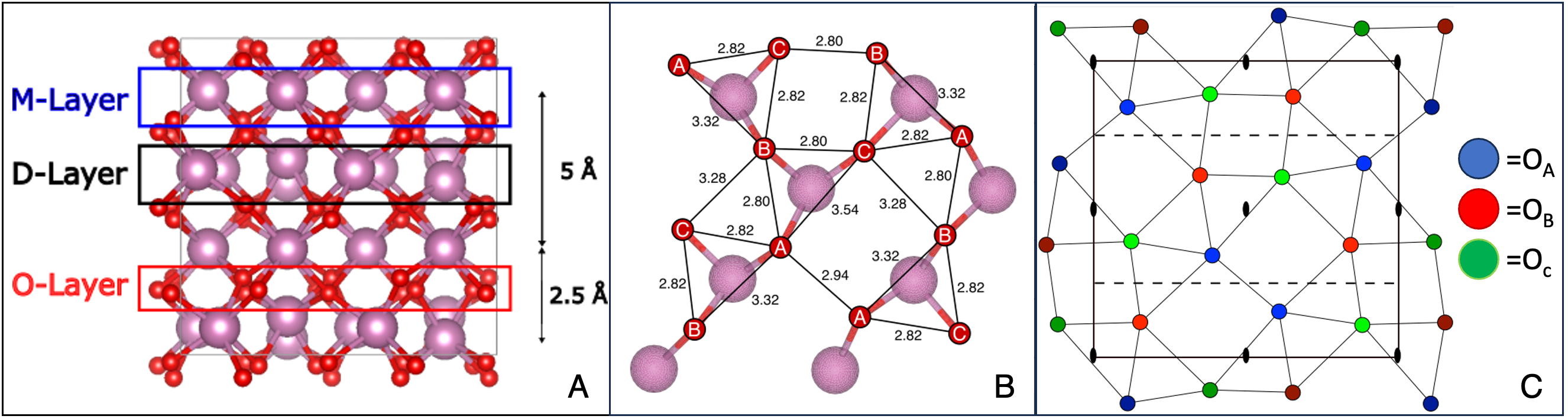}
  \caption{Structure of bulk In$_2$O$_3$ and the (001) surface. Panel A shows the three distinct layers of the In$_2$O$_3$ structure: two indium layers (M and D, where indium atoms have the same and different z-coordinates, respectively) and one oxygen layer (O layer). Equivalent indium layers are spaced 5 \r{A} apart. Panel B shows the distances between oxygen atoms at the (001) surface terminated by a MO layer. Panel C shows the nonequivalent groups on the surface, determined by using the symmetry operations in the plane of the surface (xy plane). We show three groups of inequivalent oxygen atoms: $O_A$, $O_B$  $O_C$ in blue, red and green, respectively. We show the twofold axis, whose intersection with the surface is represented as an oval,  and glide planes, displayed as dotted lines, found in the xy plane of space group Ia-3 (206).}
  \label{fig:sym}
\end{figure}

\begin{figure}
  \includegraphics[width=0.9\textwidth]{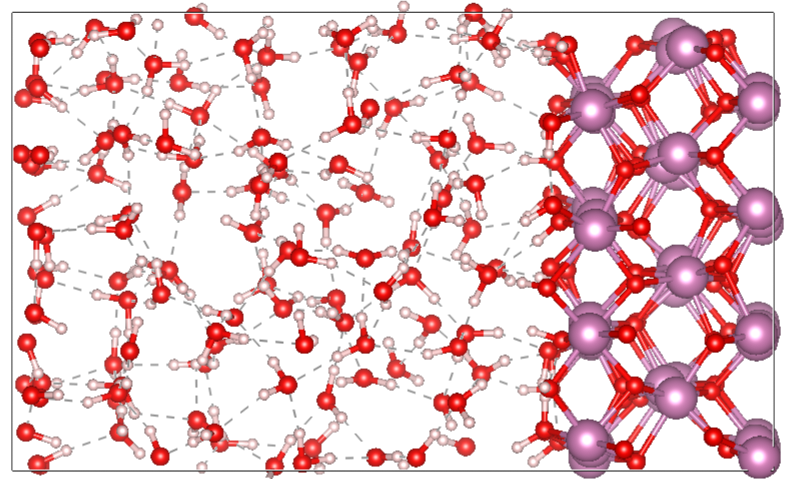}
  \caption{Structure of the (001) In$_2$O$_3$/water interface. We considered a slab of dimensions $\sqrt{2}a $ by$ \sqrt{2}a$ in the x and y directions (see SI) to increase the area of surface/water interaction; a is the computed lattice parameter of In$_2$O$_3$, 10.08 \r{A} at the SCAN level or theory, compared to the experimental value of 10.118 \r{A}. The composition of the slab shown in the figure is In$_{48}$O$_{96}$H$_{48}$ (3 layer) on top of which we placed 96 water molecules. We also considered a 5 layer slab (In$_{80}$O$_{144}$H$_{48}$) to verify the convergence of electronic structure calculations (see text). Oxygen, indium and hydrogen atoms are shown as red, purple and white spheres, respectively.}
  \label{fig:setup}
\end{figure}

Fig. \ref{fig:sym} shows the structure of bulk indium oxide, with details in the SI. We modeled the (001) In$_2$O$_3$ surface in contact with water and vacuum using symmetrical slabs, shown in Fig. \ref{fig:setup} for the solvated, hydroxylated surface (see SI). 

The hydroxyl termination was chosen as it enables the study of an important intermediate structure in the water oxidation reaction\cite{Zhu2018, Viswanathan2015} and it can stabilize the Tasker type III, polar (001) surface; indeed, existing theoretical results\cite{Zhou2008, Agoston2011} and  experimental evidence on commercial samples suggest some degree of hydroxylation of the surface upon water exposure \cite{Purvis2000}.  

 All first principles molecular dynamics simulations were carried out using the Qbox code\cite{Gygi2008} and the SCAN meta-GGA functional\cite{Sun2015}. Electronic properties of selected snapshots were investigated with the non-empirical Range-Separated Hybrid Functional SE-RSH \cite{Zhan2023}. In our calculations we used optimized norm-conserving Vanderbilt (ONCV) pseudopotentials\cite{Schlipf2015}, and a plane wave basis set with a kinetic energy cutoff of  65 Ry. The $\Gamma$ point was used to sample the Brillouin zone of the periodically repeated supercells representing the interface. All simulations were carried out with a time step of 20 au (0.48 fs), at a target temperature of 330 K, to approximately correct for some of the inaccuracies of the SCAN functional for water at room temperature; a variation of 30 K close to room temperature is expected to be  inconsequential for indium oxide, whose Debye temperature has been reported between 420 \cite{Bachmann1981} and 811 K \cite{Preissler2013}. 

NPT simulations were performed to optimize the supercell dimensions for the solvated system, with the z cell dimension (z being the direction perpendicular to the surface) considered equilibrated when the computed density of water was approximately the same as obtained with the SCAN functional in bulk water (1.05 g/cm$^3$). The optimized parameters of the 3 layer slab are: 14.26\r{A} by 14.26\r{A} by 23.67\r{A}. 

Production NVT simulations were run for $\sim$ 15 (6) ps for 3 (5) layer slabs interfaced with vacuum. Following an $\sim$ 8 ps NPT equilibration, $\sim$ 40 ps simulations in the  NVT ensemble were then performed on aqueous interfaces. From these trajectories, structural properties, including the OH tilt angle, mass density, and vibrational density of states were obtained. Infrared spectra were computed using the Fourier transform of the dipole-dipole autocorrelation function sampled on a limited portion of the trajectory, amounting to $\sim$ 10 ps for both the dry surface and  interface with water. Dipoles were computed every 40 au for the dry system and every 60 au for the interface with water, to efficiently sample the OH stretching mode. 

\subsubsection{Finite Size Effects}
We investigated the dependence of the vacuum-aligned valence band maximum (VBM) and conduction band minimum (CBM) and the band gap of In$_2$O$_3$ as a function of the number of indium layers, as computed at the SCAN and hybrid levels of theory (see Table S2). In our calculations we considered an odd  number of layers, representing symmetrically terminated slabs, to account for the alternating In layers present in the bulk material. The values shown in the table are obtained in the absence of water, with  the surface fully terminated by hydroxyls. In the case of SCAN calculations with 3 and 5 layers, eigenvalues were obtained by averaging over 40 configurations. In the case of the 7 and 9 layer systems, a short ($\sim$ 500 fs) NVT run was performed, after which a single snapshot was considered. Although SCAN results are not expected to be quantitatively accurate to describe the electronic properties of the surface and the interface, they are useful to investigate size effects, and show approximate convergence of the VBM, CBM and fundamental gap with the 5L slab, while the 3 layer slab shows sizable quantum confinement effects.  We also computed the same quantities with the SE-RSH functional and 3L and 5L layers. 
We obtained good agreement with experiments for the fundamental gap of In$_2$O$_3$ computed with the 5L slab and the SE-RSH functional (2.7  versus 2.9 eV reported at 9K \cite{King2009}).

\begin{figure}[h]
    \centering
    \includegraphics[width=0.5\linewidth]{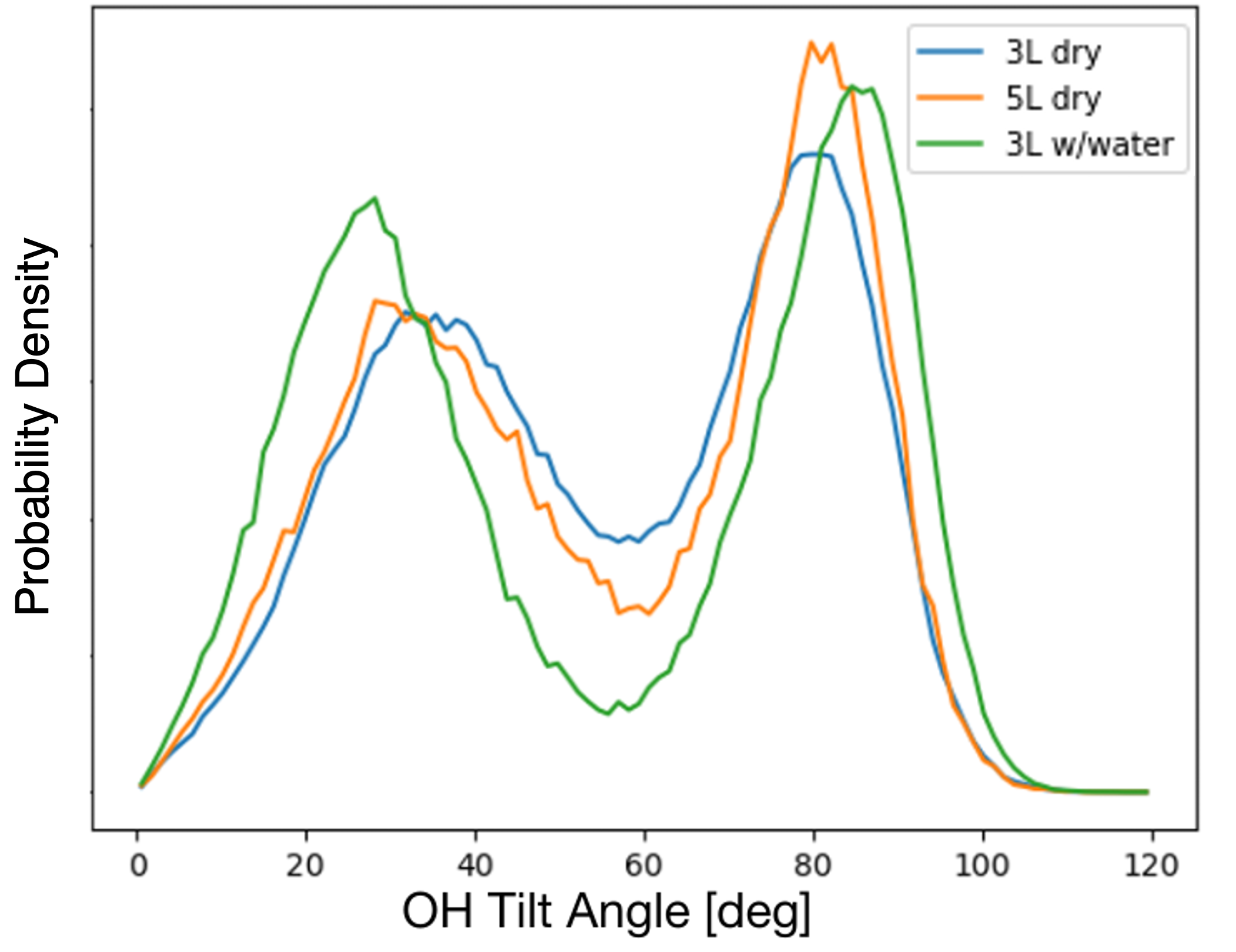}
    \caption{Computed OH tilt angles for 3 layer and 5 layer slabs representing the dry (001) In$_2$O$_3$ surface, and  3 layer slab interfaced with water.}
    \label{fig:3v5vwet}
\end{figure}

 While the electronic properties require at least 5 layers to converge, the structural properties appear to be well captured with 3 layers. Fig. \ref{fig:3v5vwet} shows the OH tilt angles at the surface computed with 3 and 5 layer slabs, showing similar results. Hence, all structural properties discussed below are reported for the 3 layer systems, while all electronic properties (band alignments) are computed for 5 layer slabs, at the SE-RSH level, on SCAN trajectories.

\section{Results}
\subsection{Dry Surface}

We first discuss our results for the dry (001) surface of In$_2$O$_3$, which has two possible terminations:  a layer of In atoms with either the same z coordinate (M layer),  or different z coordinates (D layer, see Fig. \ref{fig:sym}). We computed the total energy of each surface in vacuum by optimizing the geometry of a slab with 5 layers and found the M layer termination to be lower in energy by 1.89 eV. 
The energy difference between M and D terminated surfaces explains, at least in part,  the wide terraces with stepped edges spaced 5 \r{A} apart, observed by Hagleitner et al \cite{Hagleitner2012}, and the terraces are expected to be exclusively terminated by the M layer. 
Given the large difference in stability between M and D terminated surfaces, all MD simulations were performed with M layers exposed to water. 

\begin{figure}[H]
  \includegraphics[width=\textwidth]{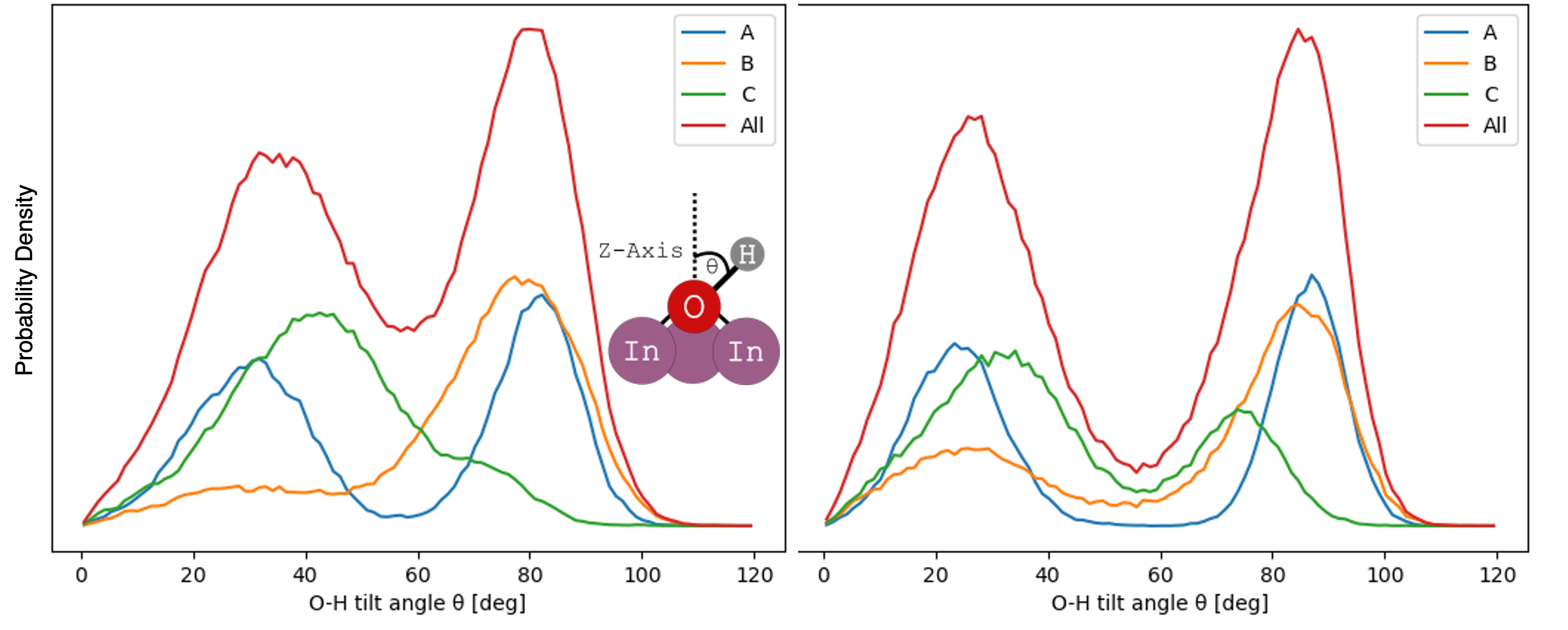}
  \caption{Angular distribution of OH tilt angles of the (001) In$_2$O$_3$/vacuum (left panel) and (001)In$_2$O$_3$/water interface (right panel). The z-axis is perpendicular to the surface. We show the total distribution (in red) as well as the contributions from inequivalent OH groups, where the varied oxygen atoms are defined in Figure \ref{fig:sym}C (see text).}
  \label{fig:OH-tilt}
\end{figure}

To characterize the environment of each hydroxyl, we plot the distribution of OH tilt angles,  as shown in Fig. \ref{fig:OH-tilt} for the three inequivalent groups of oxygen atoms shown in Fig. 1. An angle ($\theta$) of 30\degree corresponds to an out of plane hydroxyl, irrespective of whether it has a hydrogen bond acceptor interaction; $\theta$ = 60\degree corresponds to a so-called "unrestrained hydroxyl", which has no preferential orientation and can freely rotate; and $\theta$ = 90\degree to an in-plane hydroxyl, with a hydrogen bond donor interaction. 
The angular distribution of the $O_A-H$ site is bimodal with two peaks at 30\degree and 90\degree, while the distribution of $O_B-H$ sites has a main peak at 80\degree. The $O_C-H$ distribution has two peaks, a main one centered at 45\degree and a weaker one  centered at 70\degree. 

From our simulations, we observe that $O_A-H$ hydroxyl groups preferentially interact with each other, forming hydrogen bonding dyads. The absence of any $O_A-H$ tilt angles around 60\degree is indicative of the lack of unrestrained $O_A-H$ hydroxyls, indeed suggesting that $O_A-H$  bind with one another, finding a stable energy minimum. $O_B-H$ groups preferentially lay in the plane of the surface, though the large tail in their angular distribution is indicative of flexibility in their configuration. Instead, $O_C-H$ groups preferentially give rise to out of plane configurations, with a smaller subset being unrestrained hydroxyls. While in plane $O_C-H$ and out of plane $O_B-H$ will occasionally hydrogen bond with one another, these bonds are transient. 


\subsubsection{Vibrational Properties}
\begin{figure}[h!]
  \includegraphics[width=\textwidth]{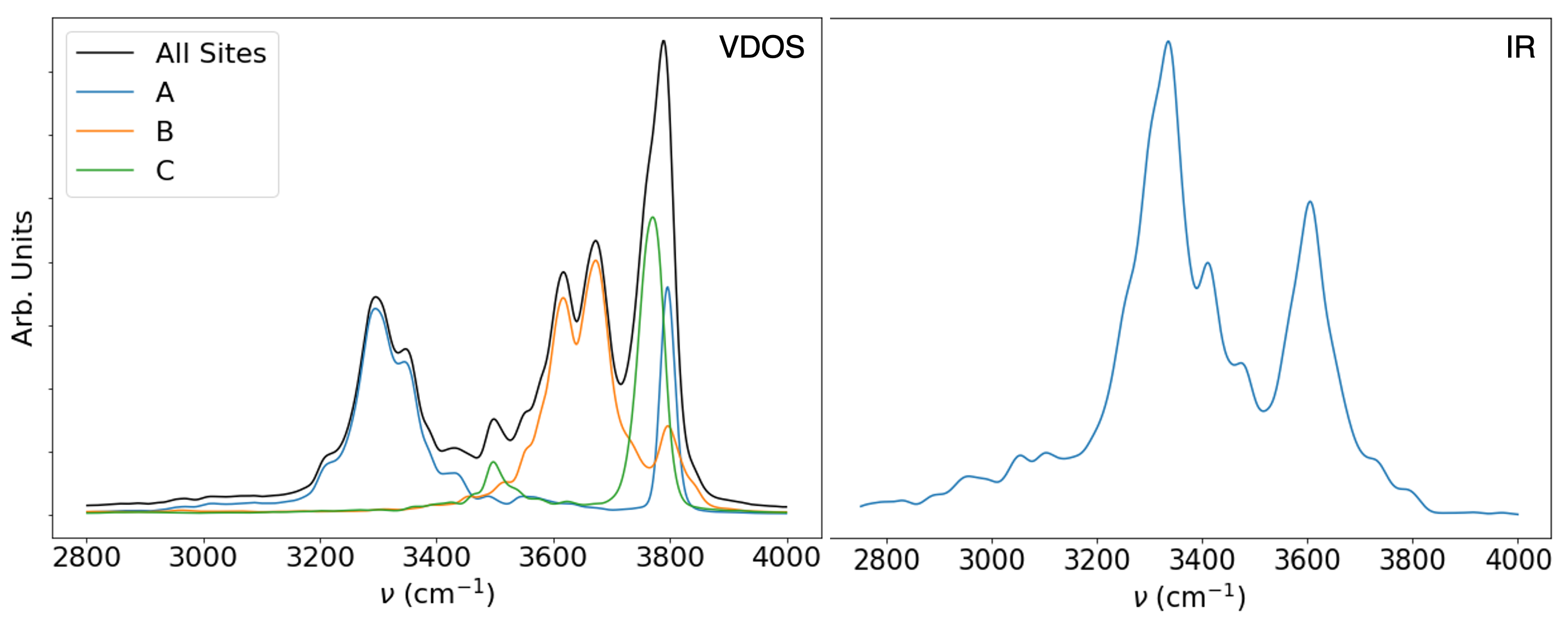}
  \caption{Vibrational density of states (VDOS) (left panel) and computed IR spectra (right panel) of the (001) In$_2$O$_3$ surface. The VDOS is decomposed into contributions from inequivalent O-H sites, where the inequivalent oxygens are defined in figure \ref{fig:sym}: $O_A$, $O_B$, and $O_C$. The IR spectrum peak at 3300 cm$^{-1}$ is assigned to hydrogen bond accepting $O_A-H$ , while the peak centered at 3600 cm$^{-1}$ is assigned to the in plane $O_B-H$ stretch (see text).}
  \label{fig:vdos-ir}
\end{figure}

In order to understand the vibrational signatures of the surface bonds identified above, we computed the surface infrared (IR) spectrum and vibrational density of states (VDOS), shown in Fig. \ref{fig:vdos-ir}. The OH stretch peak of the VDOS has been decomposed into contributions from each type of surface hydroxyl. Using  our analysis of the VDOS, we can then make peak assignments in the IR spectra. The peak at 3300 cm$^{-1}$ is assigned to hydrogen bond donating $O_A-H$ groups. These groups are in the plane of the crystal, have a strong donor interaction, and have the lowest frequencies. The peak centered at 3600 cm$^{-1}$ is primarily assigned to the in plane $O_B-H$, and the shoulder at 3700 cm$^{-1}$ to the out of plane $O_C-H$. We find that the out of plane modes have a weaker IR activity than in-plane ones, causing the shoulder at 3700 cm$^{-1}$ to be less intense. 

In order to validate our computed vibrational properties, experimental infrared spectrum were measured for (001) In$_2$O$_3$ nanocubes (see SI). The experimental spectrum has 3 peaks: a broad one at 3228 cm$^{-1}$, and twin peaks at 3650 cm$^{-1}$ and 3675 cm$^{-1}$. 
Although the agreement between theory and experiments is only qualitative, it is interesting to find that the computed IR spectrum captures the two main peaks found in experiments.

\subsection{Aqueous Interface}
We now turn to discussing the surface in contact with water. To generate aqueous interfaces, a 96 molecule water cell was interfaced with the hydroxylated surface terminated by an M layer. Initial values for the vacuum separation from the surface were assumed to be similar to that of (0001) Al$_2$O$_3$, with the first water layer at  2.3 \r{A} - 2.5 \r{A} above the surface \cite{Eng2000, Huang2014}. Further NPT simulations were carried out  to  optimize the water-surface separation, and we determined that the first water layer is approximately positioned at 1.8 \r{A} from the surface, closer than in the case of alumina (0001). 

Following NPT simulations, a single 40 ps NVT trajectory was generated to collect data on the (001) In$_2$O$_3$/water interface. Density and radial distribution functions of water are presented in the SI. From the density plots, a single interfacial water layer is identified, after which the density of water is bulk like, and attains an average density value in agreement with the results reported in Ref.\cite{LaCount2019}. 

\subsubsection{Hydroxyl Tilt Angles}
To understand the effect of water on the surface hydrogen bonding network, we plot the distribution of tilt angles for the wet surface in Fig. \ref{fig:OH-tilt}. As in the case of the dry surface, the distribution of $O_A-H$ hydroxyl groups is bimodal, with peaks at 25\degree and 87\degree. The $O_B-H$ distribution has two peaks, a main one centered at 85\degree and a less intense one centered at 25\degree. $O_C-H$ has two peaks,  centered at 30\degree and  75\degree. In each case, the peak positions are shifted, relative to the dry surface,  with a depletion of hydroxyls oriented around 60\degree . 

Such depletion and the presence of bimodal distributions are indicative of a  templating effect of water. Surface hydroxyls tend to hydrogen bond with water molecules, as either donors or acceptors. In the case of acceptors, OH groups adopt predominantly planar configurations, while  in the case of donors, they adopt out of plane configurations,  leading to the observed slight shift in peak positions, relative to the dry surface. The interaction between the first water layer and the surface likely contributes to the stabilization of the (001) surfaces in water, as observed by Agoston and Albe \cite{Agoston2011}. 

\subsubsection{Vibrational Properties}

\begin{figure}[H]
  \includegraphics[width=\textwidth]{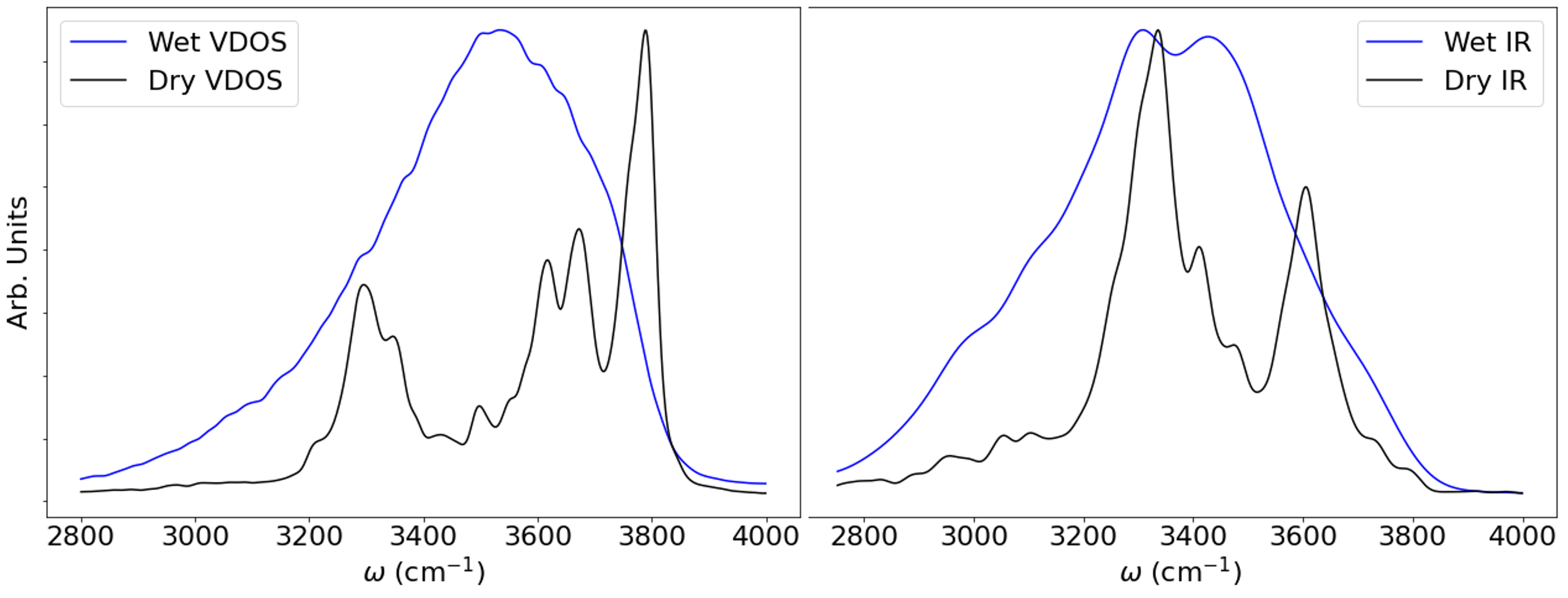}
  \caption{Vibrational Density of States (VDOS) (left panel) and computed IR spectra  (right panel) of the (001)  In$_2$O$_3$ surface interfaced with water, compared to the respective quantities obtained for the dry surface.}
  \label{fig:wet-ir}
\end{figure}

Fig. \ref{fig:wet-ir} shows the VDOS and IR spectra computed for the interface of the (001) In$_2$O$_3$ with water. As expected, the OH stretch is dominant in the IR spectrum, with the surface modes being barely resolvable. The VDOS shows a broadening of the peaks relative to the corresponding ones in the dry surface, due to the interaction of hydroxyl groups with water molecules.  The most notable change is for the $O_C-H$ group (see SI), whose out of plane stretching is red-shifted due to hydrogen bonding with the first water layer. A similar redshift is observed in the out of plane $O_A-H$ groups. These groups, which have an acceptor interaction with an in plane $O_A-H$ and a donor interaction with the first water layer, experience a larger shift, relative to the dry surface, than those $O_C-H$ groups, which do not interact as strongly with other surface hydroxyls. The peaks of the distribution of in plane $O_A-H$ are also red-shifted relative to the dry surface, by almost $\sim$ 200 cm$^{-1}$, as a consequence of hydrogen bond acceptor interactions with water. The $O_B-H$ distribution's peak, conversely, experiences only a slight shift in position, but does exhibit a large broadening.

\subsection{Electronic Properties}
\label{sec:elec-properties}
\begin{figure}[H]
  \includegraphics[width=\textwidth]{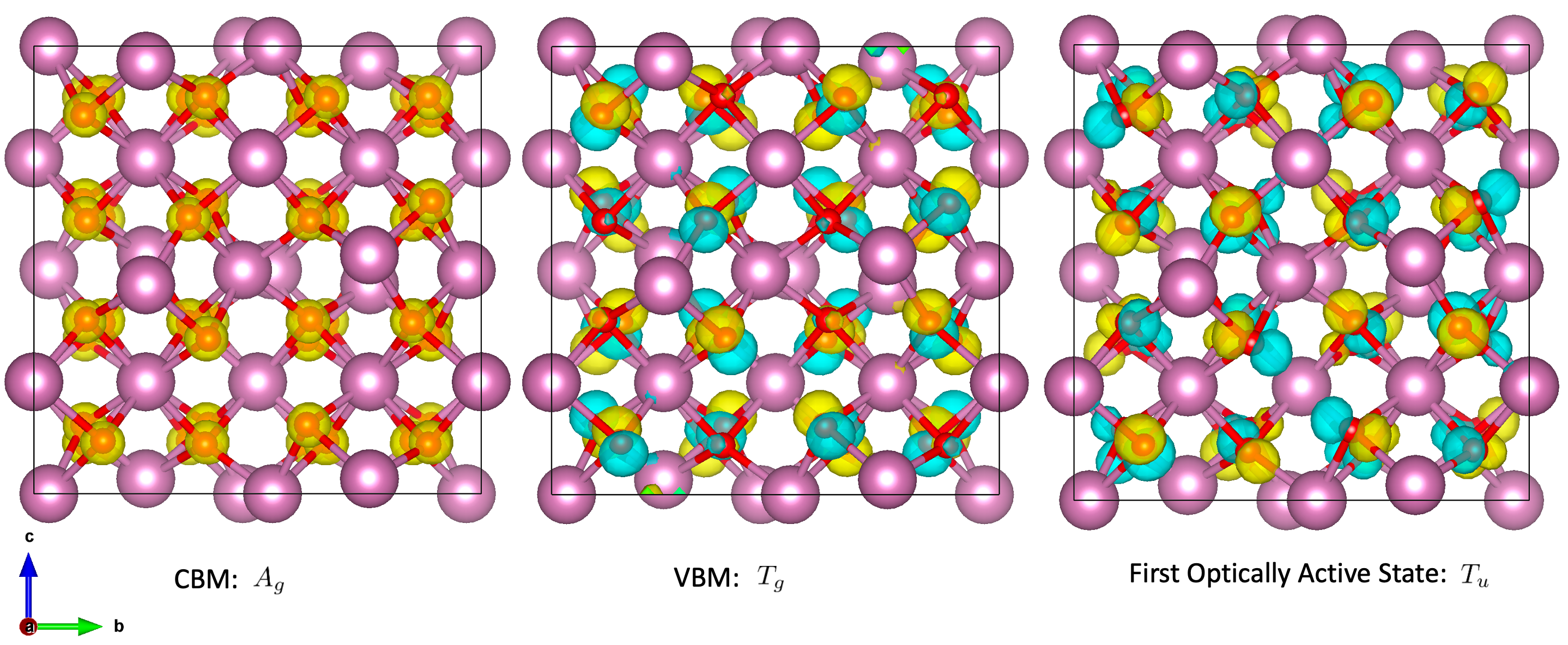}
  \caption{Square moduli of the single particle wavefunctions of the conduction band minimum (CBM), valence band maximum (VBM) and first optically active state in bulk In$_2$O$_3$. The first optically active state has the  appropriate symmetry for an allowed transition (T$_u$), as reported in \cite{Walsh2008}.}
  \label{fig:optical_sym}
\end{figure}

Having determined the structural and vibrational properties of the (001) indium oxide surface and of the aqueous interface, we now discuss their electronic properties.  We report results obtained with the screened-exchange range-separated hybrid functional (SE-RSH) \cite{Zhan2023} which has been tested on several oxides \cite{Zhan2024}, obtaining  excellent agreement with experiments for band gaps and dielectric constants.
We computed the fundamental and optical gaps of bulk In$_2$O$_3$ with a cell with 80 atoms (see Fig. \ref{fig:sym}) and of the 5L and 3L slabs at zero temperature (see Table S3). The fundamental gaps of the bulk and the 5L slab are close to each other (as observed also with the SCAN functional), 2.70 and 2.74 eV, and in good agreement with experiments (2.9 eV at 9K \cite{Bourlange2008, King2009, Irmscher2013}). Our computed optical gap for the 5L slab is also in good agreement with the UV-Vis spectra reported in Ref. \cite{King2009} for the (001) surface, 3.59 eV versus 3.55 eV. The optical gap was determined by computing the dipole-moment transitions between the conduction band and the valence states below the VBM (which is optically dark \cite{Walsh2008}) as a function of energy, and determining the onset energy for absorption (See SI).  We find that the optical gap of the bulk ($\simeq$ 4 eV) is larger than that of the 5L slab ($\simeq$ 3.55 eV), although also in the bulk the optical activity in correspondence of 3.5-4.0 eV is nonzero, albeit weak (see SI). The square moduli of the orbitals corresponding to the VBM, CBM and first occupied optically active state are reported in Fig. \ref{fig:optical_sym} for the bulk. We find that the band gap of the 3L is larger than that of the 5L one, and the quantum confinement (QC) effect almost exclusively comes from the change in position of the parabolic CBM. The position of the VBM (which is rather flat) is similar to that found for the 5L slab. 

\begin{figure}[H]
  \includegraphics[width=\textwidth]{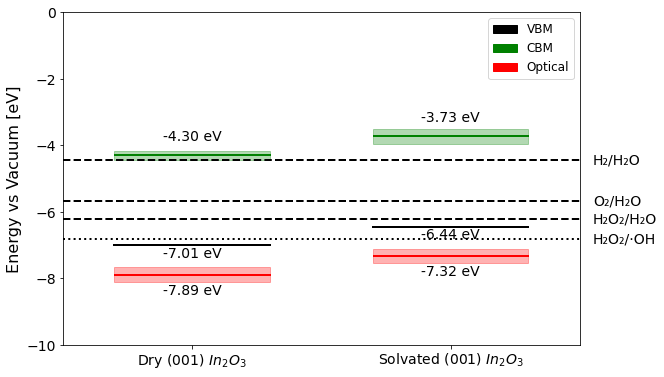}
  \caption{Alignment of In$_2$O$_3$ bands with water oxidation potentials. The bands alignment to vacuum was obtained at the hybrid level of theory (see text). As the transition from the VBM is CBM is optically dark, we show, in red, the first occupied level from which an optical transition occurs.}
  \label{fig:band-alignment-1}
\end{figure}

Fig. \ref{fig:band-alignment-1} shows the band alignment of the dry (001) surface and the surface interfaced with water, obtained with the following computational strategy. Using trajectories generated at the SCAN level of theory for the 3L slab with and without water, we determined the spread of the VBM and CBM eigenvalues due to finite temperature and the average effect of solvation (difference in the band position between the dry and solvated surface). 
We sampled our MD trajectory approximately every 300 fs for the electronic structure analysis, for a total of 60 configurations. The finite temperature spread of eigenvalues is shown by the red and green bars in Fig. \ref{fig:band-alignment-1}. The average effect of solvation was found to be $\simeq$ 0.55 eV for the VBM and CBM. We then assumed that the spread and solvation effects are approximately the same for the 5L slab and for selected configurations we computed the position of the VBM, CBM and of the occupied level at which the first optical transition occurs, using the SEH functional. Finally, we aligned the eigenvalue with the redox potential of water using the method proposed in Ref. \cite{VandeWalle1987} and described in the SI. In the case of the 3 layer slab interfaced with water, we also computed the local density of states (LDOS, shown in Fig. S10), and we found a good agreement for the position of the bands relative to vacuum, when compared to the results obtained with the method of Ref. \cite{VandeWalle1987}. 

The CBM and VBM of the surface and the highest optically active valence level straddle the redox potential of H$_2$O$_2$, with the valence states being closer to H$_2$O$_2$/H$_2$O than to O$_2$/H$_2$O, as desired for the WOR. These results show that indeed, the (001) In$_2$O$_3$ surface has promising electronic properties for the production of hydrogen peroxide. The main problem that remains to be addressed is the reduction of the large optical gap of the material. A possible strategy is to utilize thin films under tensile strain. We computed the fundamental and optical gap of the crystal under 2$\%$ tensile strain and found a reduction of 0.35 eV in both gaps, mostly due to a lowering in energy of the conduction band, which would be approximately located at -4.08 e V on the scale of Fig. \ref{fig:band-alignment-1}.  
Interestingly, our results for the electronic properties of indium oxide are consistent with and can explain recent experiments reported by Wang et al. \cite{Wang2024}. These authors observed that passivating BiVO$_4$ with an  In$_2$O$_3$ layer leads to an enhancement of the photoelectrochemical performance of the BiVO$_4$-In$_2$O$_3$ photoanode, achieving earlier photocurrent onsets and much higher photoinduced currents ($\sim$ 11 mA cm $-2$ @ 1.8 V$_{RHE}$), relative to both bare BiVO$_4$ and other heterostructures. Note that In$_2$O$_3$ is likely to be under tensile strain on BiVO$_4$, given the lattice mismatch of the two oxides. The authors of Ref. \cite{Wang2024} also remark  that the In$_2$O$_3$ coating greatly enhances the charge transfer, and promotes the WOR reaction (while suppressing the OER reaction). 
We suggest that the enhancement from  In$_2$O$_3$ observed experimentally may be further improved by coating the BiVO$_4$ surface selectively with a (001) In$_2$O$_3$ layer, as the (001) surface has superior photocatalytic properties and water stability relative to the (111) surface (Wang et al. report a mix of (222) and (004) facets for the In$_2$O$_3$ layer). 


\section{Conclusions}

 We have characterized the surface and aqueous interface of In$_2$O$_3$, a promising material for the production of H$_2$O$_2$.
 Specifically we considered the (001) surface, and using first principles simulations we investigated  the structural, vibrational and electronic properties of the dry and hydroxylated surface, and its interface with liquid water. The bulk structure of In$_2$O$_3$ is composed of alternating inequivalent layers of In, separated by an oxygen layer. Our calculations reveal the enhanced stability of a specific  indium layer on the (001) facet (the M layer in Fig.1), in contact with --OH groups. We find that the surface has a complex structure, with nonequivalent --OH sites, which we characterized using IR spectroscopy.  The presence of water modifies the structure of the --OH groups terminating in the surface, especially their mutual interaction.  Interestingly, the first water layer is found much closer to the surface in In$_2$O$_3$ when compared to other group III oxides. \cite{Eng2000, Huang2014}.
Our structural and vibrational analyses have been carried out at the DFT/SCAN level of theory. We investigated the electronic properties of the surface and interface using a hybrid density functional that has shown great promise in reproducing the band gaps of several oxides, and has the ability to correctly reproduce the variation of the dielectric properties of the oxide interfaced with water. We find good agreement with experiments for the fundamental and optical gap of the bulk and surface. Most importantly, our calculations show that the CBM and VBM of the surface and the highest optically active valence level straddle the redox potential of H$_2$O$_2$, with the valence states being closer to H$_2$O$_2$/H$_2$O than to O$_2$/H$_2$O, as desired for the WOR. In addition, we show that by growing thin films of indium oxide under tensile strain, one may achieve a reduction of the fundamental and optical gaps, e.g. by 0.35 eV with a strain of 2$\%$, which would be beneficial to increase the range of the solar spectrum that can be absorbed by the material.  Work is in progress to identify additional strategies to further decrease the optical gap, while maintaining the favorable position of the bands.

\section{Acknowledgements}
We thank Zifan Ye and Viktor Rozsa for their help with first principles simulations and for many helpful discussions. We also thank Giacomo Melani, Karen Mulfort, Seth Darling and Adam Hock for helpful discussions. This work was partly supported as part of the Advanced Materials for Energy-Water Systems Center, an Energy Frontier Research Center funded by the US Department of Energy, Office of Science, Basic Energy Sciences, and by the computational materials science center MICCoM, and by the U.S. Department of Energy, Office of Science, Basic Energy Sciences, Materials Science and Engineering Division. We acknowledge the use of the computing resources provided by the University of Chicago Research Computing center and by NERSC.

\section{References}
\bibliography{bib}


\end{document}


\section{Properties of photoanodes for water splitting}

We report below several measured properties of promising photoanodes for water splitting (Table S1) and some of the properties of the (001) In$_2$O$_3$ surface  measured in ultra high vacuum (Table S2).

\begin{table}[H]  
    \centering
    \begin{adjustbox}{max width=\textwidth} 
        \begin{tabular}{llllll}
            \hline
            Material & Peak FE & Onset V & J@ 1.8 V$_{RHE}$ & H$_2$O$_2$ accumulation  & Electrolyte \\
            &(@V$_{RHE}$) & (V) & (mA cm$^{-2}$) & (mol min$^{-1}$ cm$^{-2}$) \\
            \hline
            WO$_3$ \cite{Dias2016}& N/A & 0.4 & 2.6 (65 \degree C)& N/A & 3 M MsOH \\
            WO$_3$ \cite{Dias2016} & N/A & 0.7 & 1.3(25 \degree C)&N/A & 3 M MsOH \\
            SnO$_{2-x}$ \cite{Fan2013}& N/A & N/A & 10$^{-5}$ & N/A & 1 M Na$_2$SO$_4$\\
            BiVO$_4$ \cite{Shi2017}& 95\%@2  & N/A & N/A & N/A& KHCO$_3$  \\
            BiVO$_4$ \cite{Baek2019}& 90\%@2.4 & 0.7 & 2.25& 2 @ 3 V$_{RHE}$ & 2 M KHCO$_3$ \\
            WO$_3$/BiVO$_4$ \cite{Fuku2016}& 54\% @ 1.5  & 0.4 V &3.7 & $\sim$ 0.35 1.5 V$_{RHE}$&2 M KHCO$_3$ \\
            WO$_3$/BiVO$_4$/Al$_2$O$_3$ \cite{Fuku2017} & 80\% @ 1.5 &  0.5 V & 4.0 & $\sim$ 1.0 1.5 V$_{RHE}$&2 M KHCO$_3$ \\
            Gd-Doped BiVO$_4$ \cite{Baek2019}& 99\% @ 2 & 0.6 & 2.75& $\sim$ 2.5 @ 3 V$_{RHE}$& 2 M KHCO$_3$ \\
            Mo-Doped BiVO$_4$ \cite{Jeon2020}& 40\% @ 2 & 0.5 & 2.0& $\sim$ 0.66 @ 3 V$_{RHE}$& 1 M KHCO$_3$ \\
            FeO(OH)/BiVO$_4$/FTO \cite{Mase2016}& 70\% & 0.3 & N/A &0.2 &HClO$_4$ (pH 1.3) \\
            ZnO/BiVO$_4$ \cite{Qu2023} &40\% @1.5&0.6 &2.0 &$\sim$ 0.25 @ 1.5 V$_{RHE}$& 2 M KHCO$_3$ \\
            SnO$_{2-x}$ on BiVO$_4$ \cite{Zhang2020}& 86\% @1.23& 0.4 &5.2 &  $\sim$ 0.825 @ 1.23 V$_{RHE}$&1 M NaHCO$_3$ \\
            \hline
        \end{tabular}
    \end{adjustbox}
    \caption{Summary of materials used as photoanodes for the photoelectrochemical (PEC) production of H$_2$O$_2$. The faradaic efficiency (FE), onset potential (V), photocurrent density at 1.8 V vs Reversible Hydrogen Electrode (V$_{RHE}$)[a pH dependent reference electrode (E$_{\text{RHE}}$ = E$_{\text{SHE}} + 0.059 \times \text{pH}$) where SHE is the standard hydrogen electrode], H$_2$O$_2$ accumulation rate as function of the applied voltage, and electrolyte used are reported. In some cases, photocurrent density at 1.8 V$_{RHE}$ is not given in the respective papers. The electrolyte MsOH denotes CH$_3$SO$_3$H. }
    \label{tab:tab}
\end{table}

\begin{table}[H]
    \centering
    \begin{adjustbox}{max width=\textwidth, center}
    \begin{tabular}{llll}
        \hline
        Sample & Growth & Characterization  & Reconstruction  \\
        \hline
         (001) Thin Film\cite{Bourlange2009} & MBE  & AFM, XPS, TEM & Faceting, M-O$_2$, M-O-O-M \\
         (001) Thin Film \cite{Morales2009}& MBE  & MBE, STM, LEED & M-O$_2$, M-O-O-M, M-OH, M-OH$_2$\\
         (001) Films\cite{KOROTCENKOV2005}& SP & XRD, HRTEM, AFM, XPS & M-O$_2$, M-O-O-M, M-OH, M-OH$_2$ \\
         (001) Thin Film\cite{Bierwagen2009} & MBE  & RHEED, SEM, AFM & Faceting, In-ad atoms\\
         (001) Single Crystal\cite{Hagleitner2012}& flux  & STM, STS, PES, LEED,  & N/A \\
         \hline
    \end{tabular}
    \end{adjustbox}
    \caption{Summary of (001) In$_2$O$_3$ samples grown in ultra-high vacuum (UHV) and reported in recent literature. Growth methods include  molecular beam epitaxy (MBE), spray pyrolisis (SP), and flux. Films have been characterized with atomic force microscopy (AFM), X-ray photoelectron spectroscopy (XPS), transmission electron microscopy (TEM), x-ray diffraction (XRD), high-resolution transmission electron microscopy (HRTEM), reflection high-energy electron diffraction (RHEED), scanning electron microscopy (SEM), scanning tunneling spectroscopy (STS), photoelectron spectroscopy (PES), and low-energy electron diffraction (LEED). Experiments are conducted in the absence of water. In the type of reconstruction, M refers to the indium layer reported in Fig.1 of the main text.}
    \label{tab:uhv}
\end{table}

\section{Crystal Structure of Indium Oxide}

At ambient conditions, indium oxide is a body centered cubic crystal belonging to the space group 206, with a lattice constant of 10.118 \r{A} \cite{Marezio1966} at room temperature. The unit cell, shown in Fig. 1A of the main text, contains 80 atoms. Within the lattice, the 32 indium atoms occupy two different positions: 24 In atoms occupy  distorted octahedral sites (they are named "d" sites), and 8 indium atoms occupy sites with rotational symmetry C2 or C3 (they are named "b" sites).  Each In atom is octahedrally coordinated to 6 oxygen atoms. Along the (001) axis, the crystal is composed of 3 distinct alternating layers: a layer of d In atoms (D layer), a layer with both b and d In atoms  (M layer), and a layer consisting of oxygen atoms (O layer). We note that the M layer contains atoms that have the same z-coordinate, while the D layer contains atoms with different z-coordinates. Adjacent In layers are spaced by approximately 2.5 \r{A}, while identical In layers repeat approximately every 5 \r{A}, as seen in Fig. 1B of the main text.

\section{Supercell Geometries used in Molecular Dynamics Simulations } 
\subsection{Bulk Crystal}
The construction of the slabs used in our MD simulations proceeded in the following way: 
starting  from a 80 atom bulk In$_2$O$_3$ primitive cell, we  generated a slab containing 3 In layers (3L), by removing the bottom most O-D layers, obtaining a layer sequence: [O-M-O-D-O-M-O]. We also generated  a slab containing 5 layers (5L); to do so the primitive cell was replicated in the z-direction, yielding  a layer sequence [O-D-O-M-O-D-O-M][O-D-O-M-O-D-O-M]. The bottom most O-D layers and the top most D-O-M layers were removed, leaving the sequence [O-M-O-D-O-M-O-D-O-M-O]. All surfaces were passivated by adding a hydrogen atom 0.95 \r{A} Angstrom above each terminal oxygen. 
To generate a $\sqrt{2}a$ by $\sqrt{2}a$ cell, the bulk cell (whose lattice constant is computed to be 10.08 \r{A} at the SCAN level of theory) was first replicated in the z direction. Then it was rotated by 45 degrees about the z axis. The cell axes are defined as a' = $\sqrt{2}$a ; b' = $\sqrt{2}$b ;
c' = c. Finally, atoms were folded back into the Wigner Seitz cell.

\subsection{Hydroxylated Surface}

The symmetry of the crystal is broken at the surface, and only symmetry operations in two dimensions are allowed. Using the operations found in space group 206, we identified 3 symmetry nonequivalent oxygen sites, corresponding to 3 different environments of the surface hydroxyls. These groups have been color-coded and labeled: A (blue), B (Red) and C (green) hydroxyls; the corresponding oxygen atoms (O$_A$, O$_B$ and  O$_C$) are shown in Fig. 1 of the main text.  We note the similarity of our classification to that by Golovanov et al., \cite{KOROTCENKOV2005} who classify each surface oxygen of the oxygen-terminated surface  by its distance from the In layer immediately below. They identify O$_{I}$, O$_{II}$ and O$_{III}$ which are 1.43 \r{A}, 1.31 \r{A} and 0.95 \r{A} from the In plane, respectively. These correspond to the sites named here  $O_B$, $O_A$, and $O_C$, respectively. 

In their ideal lattice configurations, the oxygen atoms at the surface form triangular  and rectangular configurations. The $O_A-O_A$ distance is 2.94 \r{A}, and these sites are found as adjacent points on a parallelogram. Adjacent $O_A-O_B$ distances range from 2.80 \r{A} to 3.32 \r{A}. Adjacent $O_A-O_C$ distances range from 2.82 \r{A} to 3.54 \r{A}. Adjacent $O_B-O_C$ distances range from 2.80 \r{A} and 2.82 \r{A} on parallelogram corners, and 2.82 \r{A} to 3.28 \r{A} on triangle vertices. The $O_A$ atoms are 1.34 \r{A} above the first In layer below, and $O_B$ and $O_C$ are 1.42 \r{A} and 0.96 \r{A} above, respectively.

Each inequivalent oxygen  atom has 3 neighbors, approximately equidistant. The neighbors of $O_A$ are:  $O_A$ (at a distance of 2.94 \r{A}),  $O_B$ (at a distance of 2.80 \r{A}) and  $O_C$ (at a distance of 2.82 \r{A}).  The neighbors of $O_B$ are: $O_A$ (2.80 \r{A}), $O_C$ (2.80 \r{A}) and $O_C$ (2.82 \r{A}). The neighbors of $O_C$  are $O_A$ (2.82 \r{A}),  $O_B$ (2.80 \r{A}) and  $O_B$ 2(.82 \r{A}).

\section{Finite Size Effects in Electronic Structure Calculations}

\begin{table}[h]
    \centering
    \begin{tabular}{cccc}
    \hline
        Number of Layers  &VBM$_{vac}$& CBM$_{vac}$& E$_g$\\
        \hline
        SCAN& & & \\
        \hline
         3 &-3.03$\pm$0.24&-1.11$\pm$0.22 & 1.92$\pm$0.07\\
         5& -4.53$\pm$0.24&-3.27$\pm$0.17 & 1.28$\pm$0.10\\
         7& -4.64&-3.29 &1.35 \\
         9& -4.75&-3.44& 1.31\\
         \hline
         Hybrid& & & \\
         \hline
         3 & -6.99 & -3.59 & 3.40 \\
         5 & -6.96 & -4.26 & 2.71 \\
         \hline
    \end{tabular}
    \caption{ Vacuum aligned Valence band Maximum (VBM) and Conduction Band Minimum (CBM), and band gap (E$_g$) for 3, 5, 7, and 9 layer (001) hydroxylated In$_2$O$_3$ slabs obtained at the SCAN \cite{Sun2015} and hybrid SE-RSH \cite{Zhan2023} level of theory.  All values are in eV}
    \label{tab:finite}
\end{table}

 Table \ref{tab:finite} shows the vacuum aligned valence band maximum (VBM$_vac$) and conduction band minimum (CBM$_vac$) for 3, 5, 7, and 9 layer slabs computed at the SCAN level of theory, as well as the same quantities  computed at the hybrid level for the 3 and 5 layer slabs. While the electronic properties require at least 5 layers to converge, the structural properties appear to be well captured with 3 layers. Hence, all structural properties discussed in the main text are reported for the 3 layer systems, while all electronic properties (band alignments) are computed for 5 layer slabs.
 
\section {Experiments on indium oxide nanocubes}
\label{sec:exp-methods}
\subsection{Synthesis of cubic In$_2$O$_3$ nanoparticles.} The following synthetic procedure was adopted to obtain In$_2$O$_3$ nanoparticles: 2.031 g of potassium oleate (KOL) was added into 6 mL of oleic acid in 50 mL three-neck flask and heated to 120 \degree C with stirring, see Fig. S1. After a homogeneous solution formed, 4 mL of octadecene was added and the solution was heated to 160 \degree C. Then, 1.168 g of Indium (III) acetate was added into the solution and heated under nitrogen purge for one hour to get a transparent yellow precursor solution. 6.5 mL of oleyl alcohol was added to a new 50 mL three-neck flask and heated to 130 \degree C under vacuum for 20 min then further heated to 290 \degree C under flowing nitrogen. Next, the precursor solution was transferred to a 50 mL glass syringe and slowly injected into the new flask. After injection of the reactants, temperature was maintained at 290 \degree C for another 20 min before cooling to room temperature. Finally, the product was solvent cleaned with the use of hexane, reagent alcohol, and deionized water. Centrifugation was further applied to separate the nanoparticles and the solution. SEM imaging (IT800 Ultrahigh Resolution Field Emission SEM from JEOL and SmartSEM from ZEISS) was used to identify the morphology of each In$_2$O$_3$.

\subsection{Fourier-transform infrared spectroscopy (FTIR)} Infrared spectra were obtained using a Bruker Tensor 27 FTIR spectrophotometer equipped with a diffuse reflectance infrared Fourier transform (DRIFTS) accessory (Praying Mantis Diffuse Reflection Accessory from Harrick Science). A liquid nitrogen (LN$_2$)-cooled mercury cadmium telluride (MCT) detector was used to collect spectra from 4000-600 cm$^{-1}$, with 256 scans at a resolution of 4 cm$^{-1}$. Background spectra were acquired on granulated KBr powder at 50 \degree C, which had been dried at 250 \degree C under N$_2$ of 5 L/min flow rate overnight. The samples spectra were taken by spreading the powder on the top of KBr after cooled down from the annealing pretreatment. Data was acquired after the system was stabilized at target temperature for at least 1 hour under vacuum and constant N$_2$ purge with a flow rate of 5 L/min.

\section{Growth of In$_2$O$_3$ Nanoparticles}
\begin{figure}[H]
    \centering
    \includegraphics[width=1\linewidth]{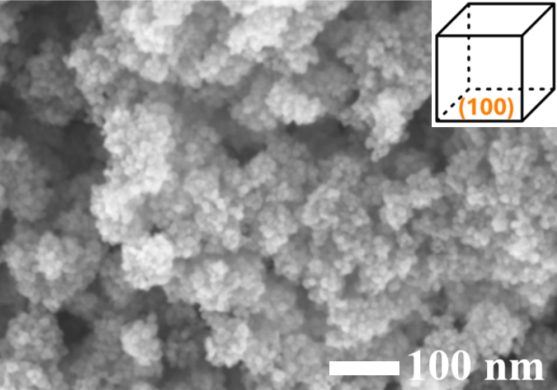}
    \caption{Cubic In$_2$O$_3$ nanoparticles, grown as detailed in section 2.2 of the main text.}
    \label{fig:nano}
\end{figure}

We report below the SEM images of cubic In$_2$O$_3$ nanoparticles grown as described in the main text (Fig. \ref{fig:nano})

\section {Measured IR spectra}

A detailed analysis of the experimental spectrum at frequencies lower than $\sim$ 2950 cm$^{-1}$ showed the presence of organics at the surface, specifically the isopropanol used to clean the surface, hence it is not surprising that the experimental and theoretical spectra only show qualitative agreement.

We note that the experimental spectrum of (001) In$_2$O$_3$ nanocubes measured here differs from published spectra of other In$_2$O$_3$ surfaces. Both IR taken of commercial samples \cite{Purvis2000} and (222) oriented nanoparticles \cite{Jothibas2015, Shen2021} contain single intense peaks in the OH stretching region, centered around $\sim$ 3400 cm$^{-1}$. The double peaks, found at 3228 cm$^{-1}$ and 3650 cm$^{-1}$ in our spectra are, therefore, indicative of hydroxylated (001) surfaces, and offer an experimental signature for determining the surface termination.

\begin{figure}
  \includegraphics[width=\textwidth]{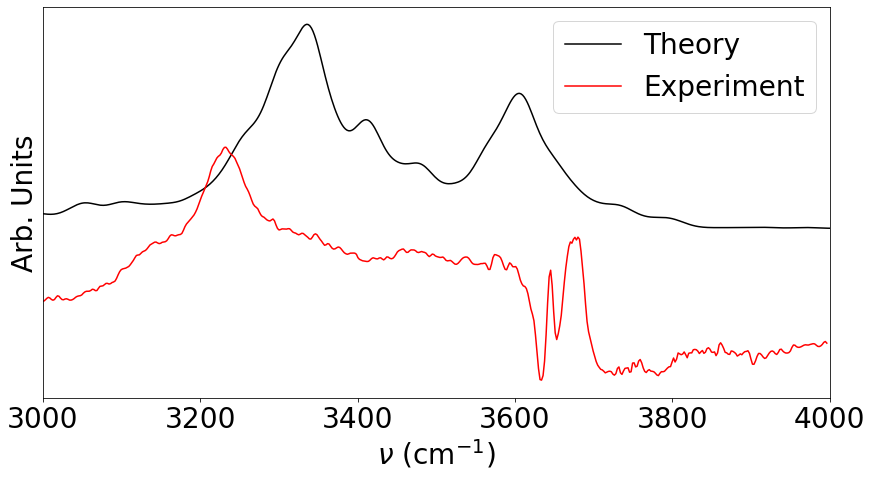}
  \caption{Comparison of experimental and calculated IR Spectra for the dry (001) surface (see text).}
  \label{fig:exp-ir}
\end{figure}
\section{Structural Properties of Liquid Water}
\begin{figure}[h]
    \centering
    \includegraphics[width=1\linewidth]{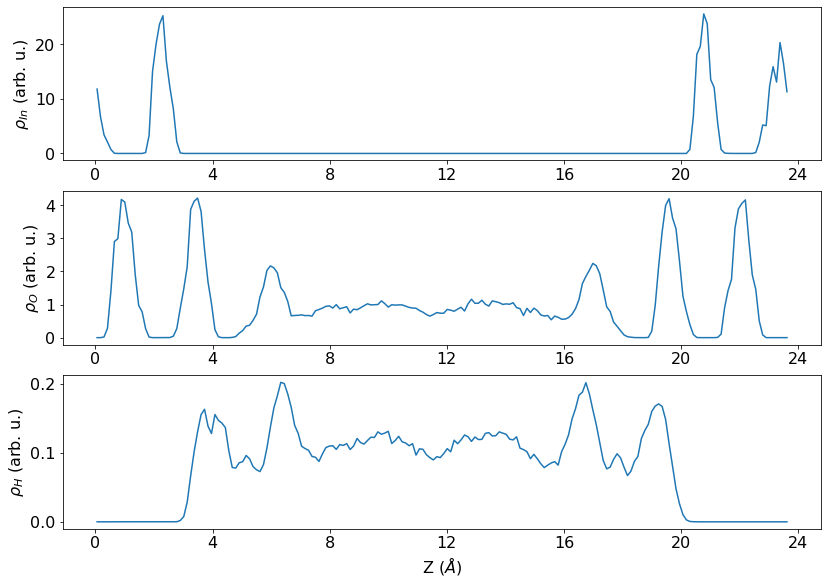}
    \caption{Atom density ( $\rho$ ) for In (upper panel), O (middle panel) and H atoms (lower panel) for a three layer slab of indium oxide in contact with water. The density of water in the slab is $\sim$ 1.05 g/cm$^3$.}
    \label{fig:den}
\end{figure}

We analyze the structural properties of liquid water in contact with the (001) In$_2$O$_3$ interface by computing the average density of In, O and H atoms, shown in Fig. \ref{fig:den}. The density of water is $\sim$ 1.05 g/cm$^3$, in agreement with the average value found for bulk water at the scan level of theory. \cite{LaCount2019}. 

\begin{figure}[H]
    \centering
    \includegraphics[width=.8\linewidth]{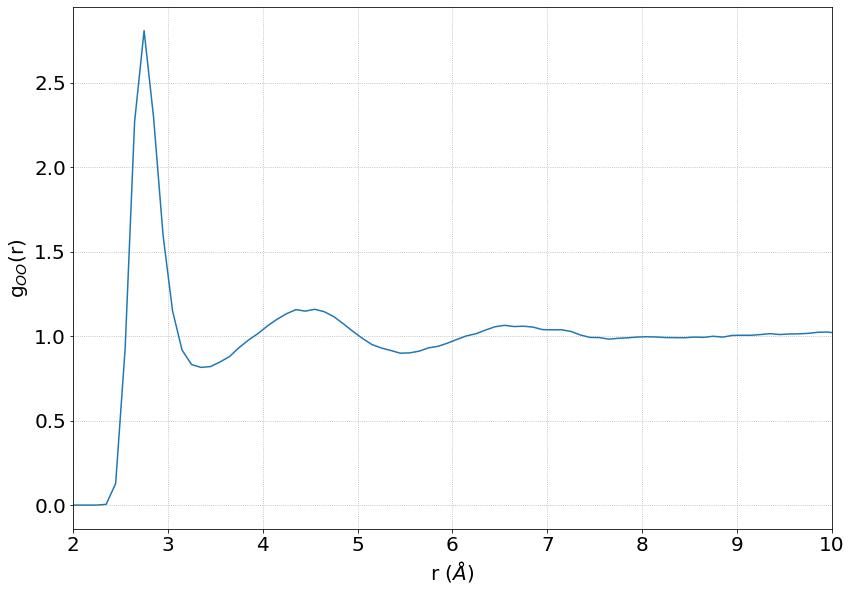}
    \caption{Radial Distribution Function for O-O liquid water in contact with the (001) surface of In$_2$O$_3$.}
    \label{fig:goor}
\end{figure}

\begin{figure}[H]
    \centering
    \includegraphics[width=.8\linewidth]{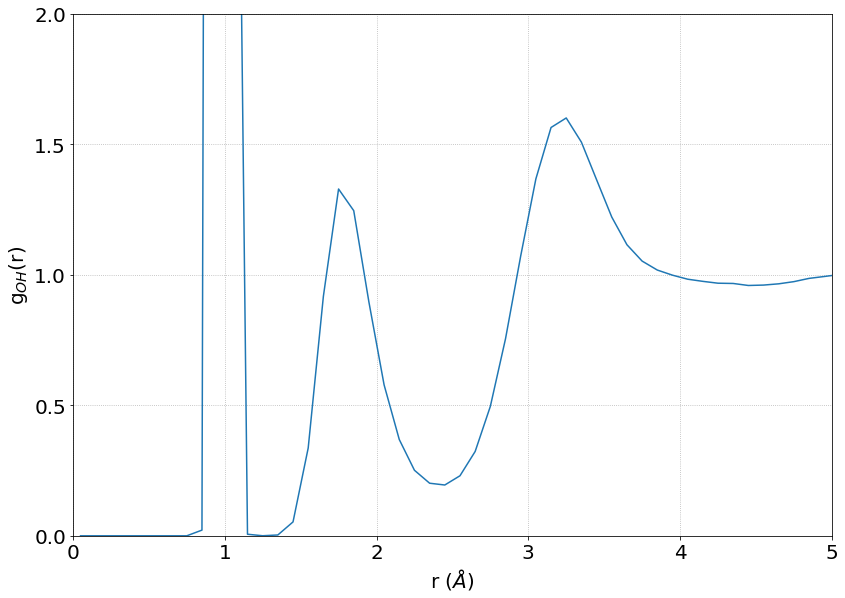}
    \caption{Radial Distribution Function for O-H of liquid water in contact with the (001) surface of In$_2$O$_3$.}
    \label{fig:gohr}
\end{figure}

\begin{figure}[H]
    \centering
    \includegraphics[width=.8\linewidth]{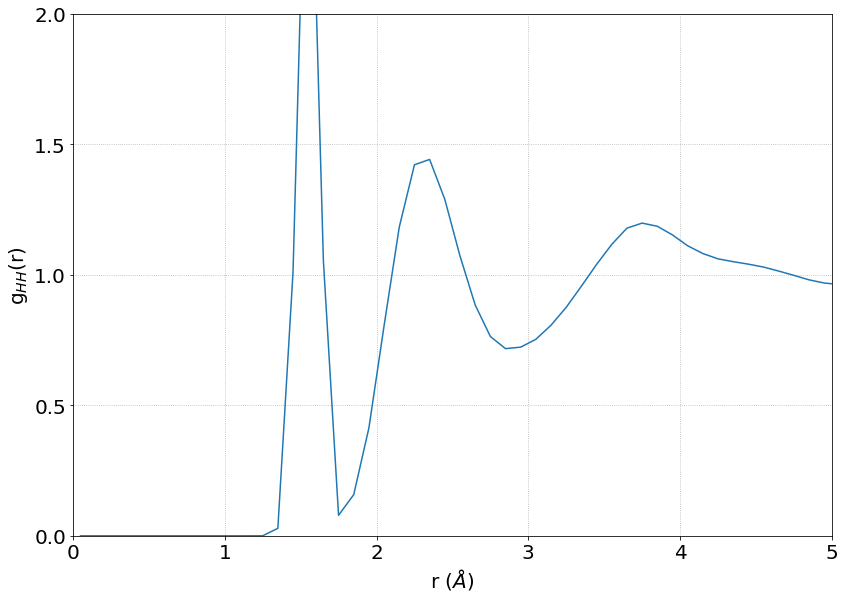}
    \caption{Radial Distribution Function for H-H of liquid water in contact with the (001) surface of In$_2$O$_3$.}
    \label{fig:ghhr}
\end{figure}

We also computed the radial pair distribution functions of the liquid. Fig. \ref{fig:goor}, \ref{fig:gohr} and \ref{fig:ghhr} shows the oxygen-oxygen, oxygen-hydrogen, and hydrogen-hydrogen pair distribution functions respectively. The O-O pair correlation function has a first peak at $\sim$ 2.75 \r{A}. The O-H pair correlation function has its first peak $\sim$ 1 \r{A} corresponding to OH bonds, a second peak at $\sim$ 1.75 \r{A} and a third peak at $\sim$ 3.25 \r{A}. The H-H pair correlation function has its first peak at $\sim$ 1.5 \r{A}, a second one at $\sim$ 2.25 \r{A} and a third peak at $\sim$ 3.75 \r{A}. In each case, the radial distribution functions agree with those previously computed for liquid water at the SCAN level of theory \cite{LaCount2019}

\section{Vibrational properties of the dry and solvated hydroxylated surface}
\subsubsection{Vibrational Properties of the Hydroxylated Dry (001) In$_2$O$_3$}
\begin{figure}[H]
    \centering
    \includegraphics[width=1\linewidth]{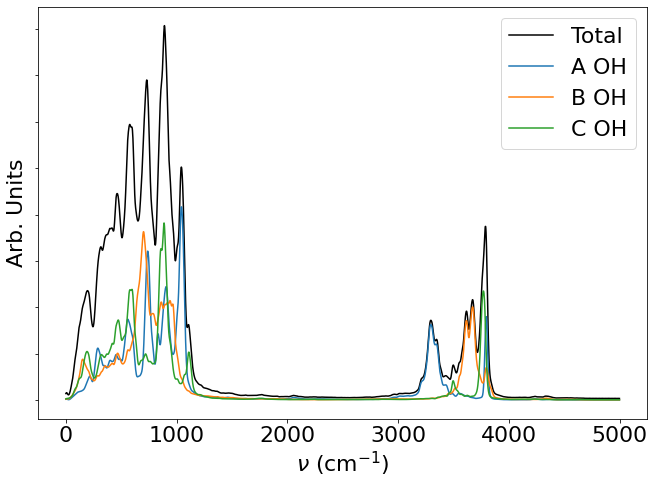}
    \caption{Total vibrational density of states  of the dry (001) In$_2$O$_3$ surface. Contributions from inequivalent O-H sites are shown. The inequivalent oxygen sites are defined in Figure 1. of the main text.}
    \label{fig:total-vdos-dry}
\end{figure}

\begin{figure}[H]
    \centering
    \includegraphics[width=1\linewidth]{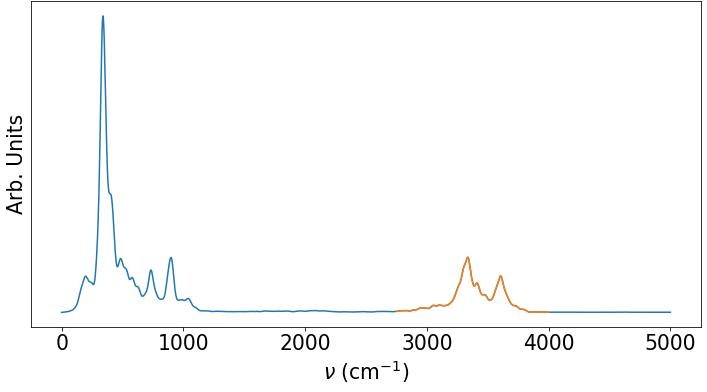}
    \caption{Infrared spectrum (IR) of the dry In$_2$O$_3$  surface. The orange part of the curve corresponds to the portion of the IR spectrum shown in the main text.}
    \label{fig:total-ir-dry}
\end{figure}

The computed vibrational density of states (VDOS) and IR spectrum of the dry (001) In$_2$O$_3$ are shown in Fig. \ref{fig:total-vdos-dry} and \ref{fig:total-ir-dry} respectively. The orange section of \ref{fig:total-ir-dry} corresponds to the portion of the IR spectrum shown in the main text. 

\subsubsection{Vibrational Properties of the (001) In$_2$O$_3$/Water Interface}
\begin{figure}[H]
    \centering
    \includegraphics[width=1\linewidth]{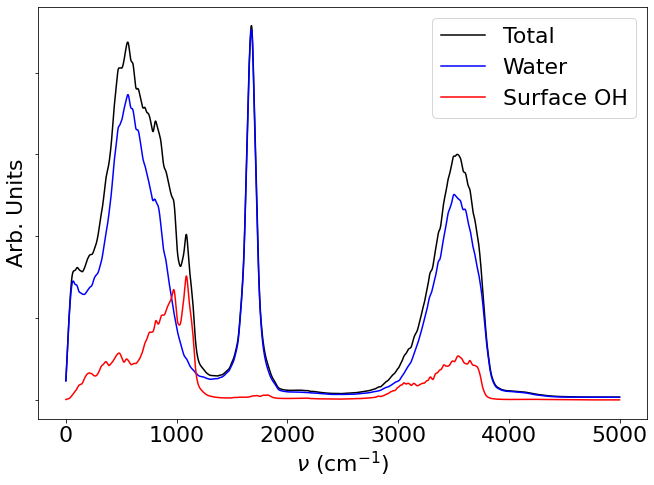}
    \caption{Total vibrational density of states  of the In$_2$O$_3$/water interface (black line) with water and surface OH contributions shown separately.}
    \label{fig:total-vdos-wet}
\end{figure}

\begin{figure}[H]
    \centering
    \includegraphics[width=1\linewidth]{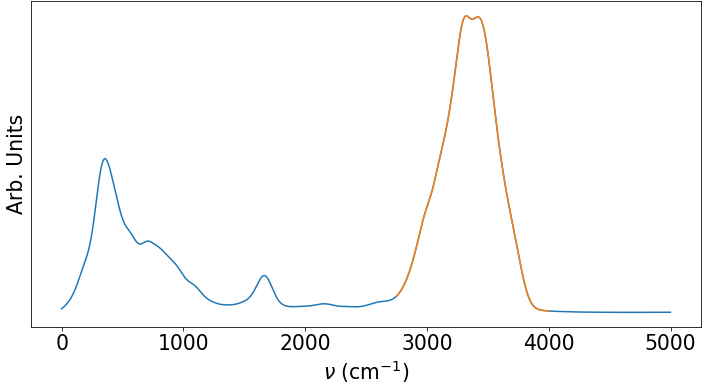}
    \caption{Infrared spectrum (IR) of the In$_2$O$_3$/water interface.}
    \label{fig:total-ir-wet}
\end{figure}

The  VDOS and IR spectrum of  the wet (001) In$_2$O$_3$ surface are shown in figure \ref{fig:total-vdos-wet} and \ref{fig:total-ir-wet} respectively. The orange section of \ref{fig:total-ir-wet} corresponds to the portion of the IR spectrum shown in the main text.

\section{Electronic Properties}
\subsection{Computation of band edges}
Band edges of indium oxide were aligned to vacuum using equation \ref{eq:1} below in order to compare their energy position with water redox potentials. 
\\
\begin{equation}\label{eq:1}
  E_{VBM} = E_{KS}^{HOMO} - \langle \tilde{V}_{bulk}\rangle - (\langle \tilde{V}_{vac}\rangle - \langle \tilde{V}_{bulk}\rangle_{slab})
\end{equation}
\begin{equation}
  E_{CBM} = E_{KS}^{LUMO} - \langle \tilde{V}_{bulk}\rangle - (\langle \tilde{V}_{vac}\rangle - \langle \tilde{V}_{bulk}\rangle_{slab})
\end{equation}
\\
Here, the KS eigenvalue is  the single particle energy of  bulk In$_2$O$_3$ while $\langle \tilde{V}_{bulk}\rangle$ is the potential computed on a bulk sample. The expression $(\langle \tilde{V}_{vac}\rangle - \langle \tilde{V}_{bulk}\rangle_{slab})$ above pertains to the dry (001) MO In$_2$O$_3$ surface. 

The effect of solvation on the electronic states was computed by calculating the difference in position of the band edges in vacuum (eq. \ref{eq:1}) and in the presence of water (eq. \ref{eq:5})
\begin{equation} \label{eq:5}
  E_{VBM}^{sol} = E_{KS}^{HOMO} - [\langle \tilde{V}_{bulk}\rangle - \langle \tilde{V}_{bulk}\rangle_{slab}] - [\langle \tilde{V}_{bulk}\rangle_{water} - \langle \tilde{V}_{int}\rangle_{water}]-\langle \tilde{V}_{vac}\rangle_{water}
\end{equation}

\begin{equation}
  E_{CBM}^{sol} = E_{KS}^{LUMO} - [\langle \tilde{V}_{bulk}\rangle - \langle \tilde{V}_{bulk}\rangle_{slab}] - [\langle \tilde{V}_{bulk}\rangle_{water} - \langle \tilde{V}_{int}\rangle_{water}]-\langle \tilde{V}_{vac}\rangle_{water}
\end{equation}
\\
$E_{KS}^{HOMO} $and $\langle \tilde{V}_{bulk}\rangle$ come from  calculations for the bulk, while $\langle \tilde{V}_{int}\rangle_{water}$ and $\langle \tilde{V}_{bulk}\rangle_{slab})$  are from calculations of the  interface  (slab). The values of $\langle \tilde{V}_{bulk}\rangle_{water}$ come from a calculation performed on a water supercell (using the same electronic parameters as the interface calculation), while $\langle \tilde{V}_{vac}\rangle_{water}$ = 3.7 eV is taken from Ref.\cite{Pham2014}.

\subsection{Local Density of States (LDOS)}
\begin{figure}[H]
  \includegraphics[width=\textwidth]{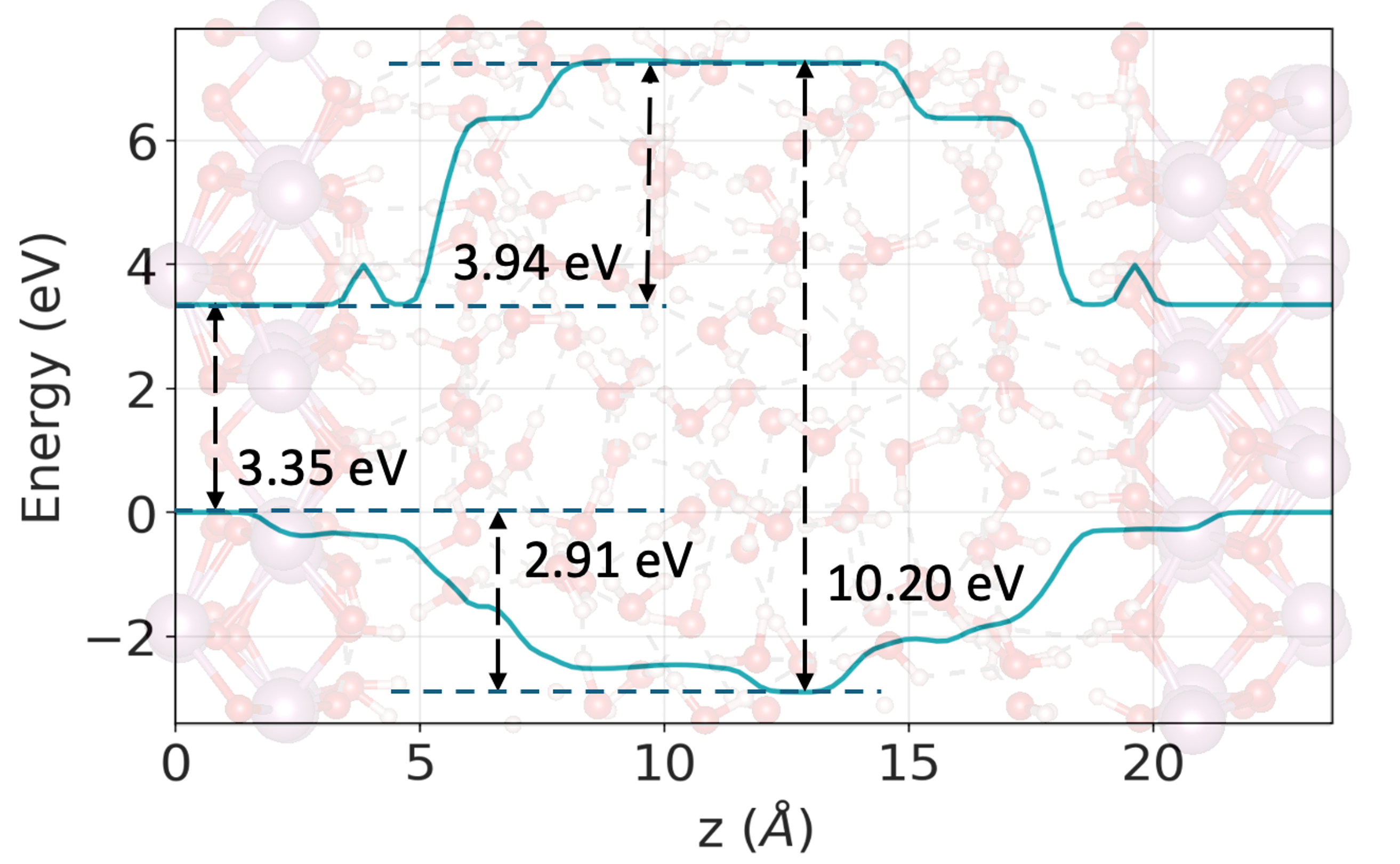}
  \caption{Local density of states (LDOS) of the In$_2$O$_3$/water interface, as obtained with the 3L slab and the SE-RSH functional. Note the band gap (3.45 eV) in close agreement with the results reported for the dry surface in Table S3 (3.4 eV). Note also that the band gap of water is in agreement with the one previously determined with the DDH functional in Ref. \cite{Gaiduk2016}. Assuming that the electron affinity of water is close to zero, as determined by SE-RSH or DDH and hence the CBM of the slab is close to the vacuum level, we find that the position of VBM and CBM of the slab are in agreement with those of Table S3.}
  \label{fig:ldos}
\end{figure}

Figure \ref{fig:ldos} shows the local density of states (LDOS) of the In$_2$O$_3$/water interface computed for the 3L slab, using the SE-RSH functional.

\subsection{Local Dielectric Constant}
\begin{figure}[H]
  \includegraphics[width=\textwidth]{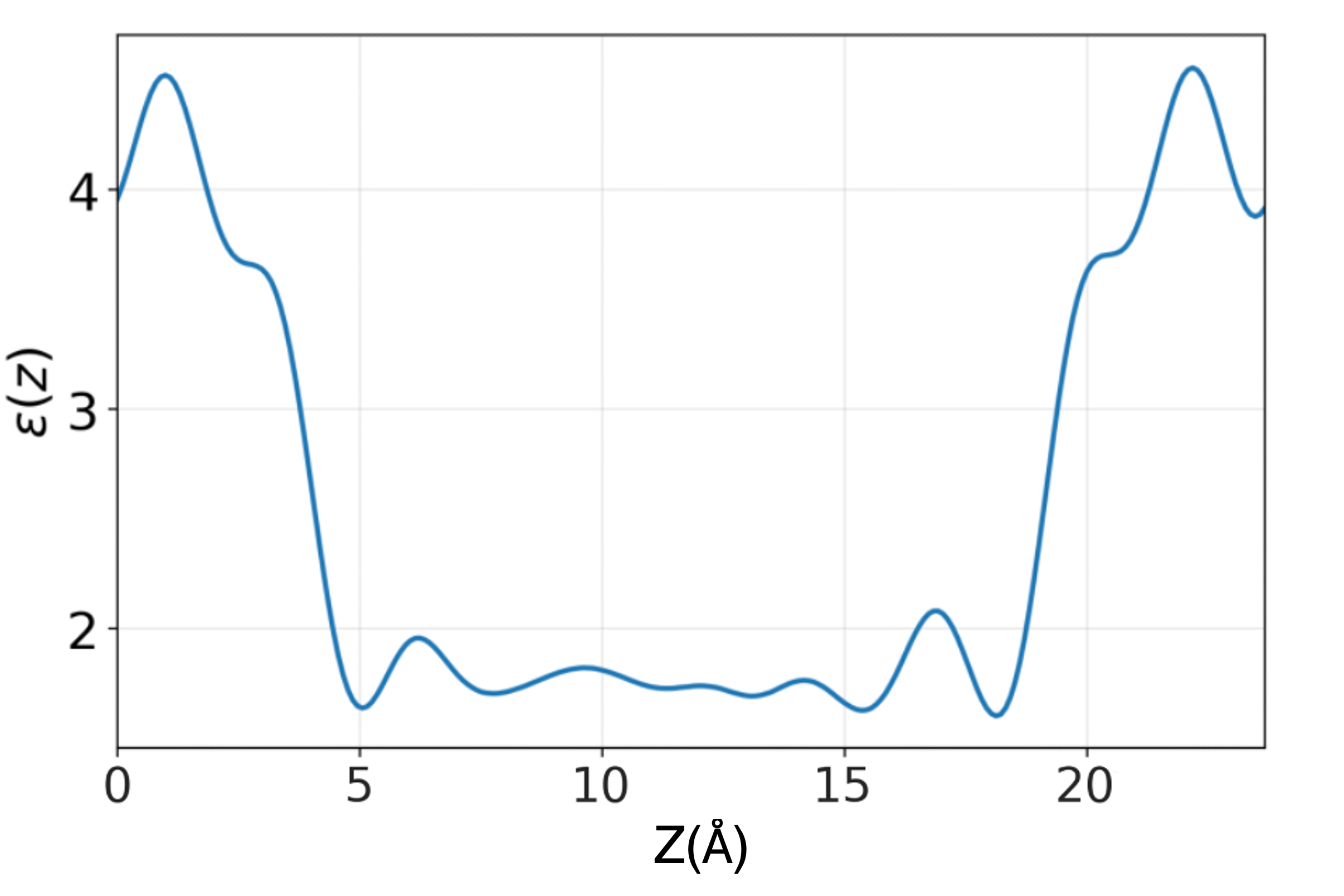}
  \caption{Value of the  high frequency dielectric constant, $\epsilon_{\infty}$, as a function of distance in the z direction perpendicular to the surface, as obtained with the SE-RSH functional.}
  \label{fig:dielectric}
\end{figure}

Fig. \ref{fig:dielectric} shows the local high frequency dielectric constant, $\epsilon_{\infty}$, as a function of distance in the z direction perpendicular to the surface for the same slab of Fig. S8. Results were obtained with the SE-RSH functional. In the In$_2$O$_3$ region we obtain a dielectric constant $\simeq$ 4.2, the same value computed for bulk In$_2$O$_3$ at the hybrid level, and close to the experimental vale of 4.0 \cite{Stokey2021}. In the water region we obtain a value of 1.75, close to the experimental value for water: 1.78.

\subsection{Dipole Transition Strengths}
\begin{figure}[H]
    \centering
    \includegraphics[height=0.9\textheight]{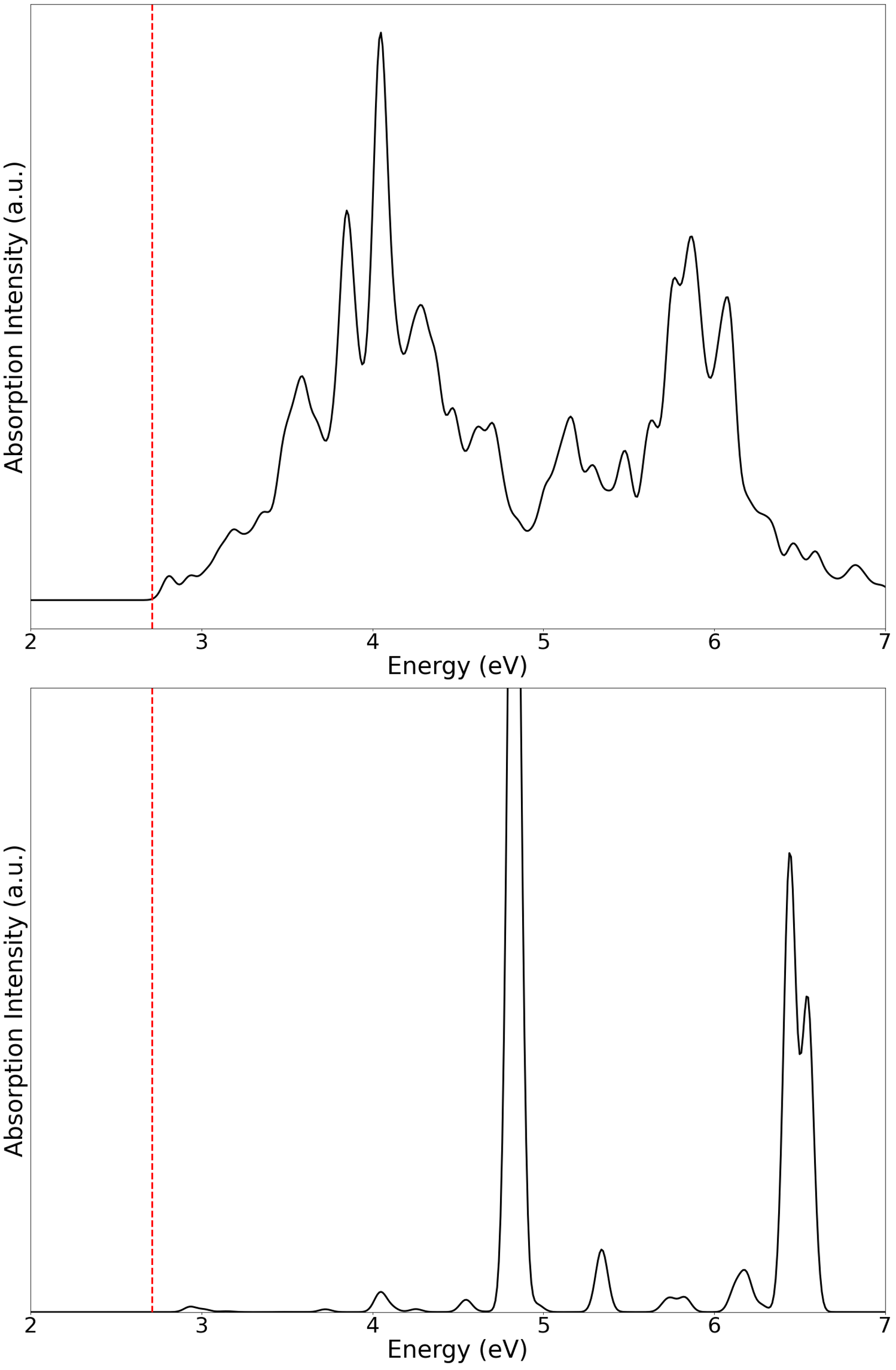}
    \caption{Dipole transition strengths from occupied orbitals to the conduction band maximum (CBM) in the 5 layer slab (top panel) and in bulk (bottom panel) In$_2$O$_3$. The transition from the VBM to the CBM is shown by a red dotted line.}
    \label{fig:dip-transition}
\end{figure}

Fig. \ref{fig:dip-transition} shows the intensity of the dipole transitions between occupied states and the CBM for both  the bulk and the 5 layer In$_2$O$_3$ slab in the absence of water. 

\section{References}

\bibliography{bib}